\documentclass[journal]{IEEEtran}

\usepackage{cite}
\usepackage{amsmath,amssymb,amsfonts}
\usepackage{algorithmic}
\usepackage{graphicx}
\usepackage{textcomp}
\usepackage{xcolor}
\usepackage{booktabs}
\begin{document}

\title{PuppetAI: A Customizable Platform for Designing Tactile-Rich Affective Robot Interaction}


\author{Jiaye~Li,
        Tongshun~Chen,
        Siyi~Ma,
        Elizabeth~Churchill,
        and Ke~Wu 
\thanks{J. Li, and E. Churchill are with the Department of Human-Computer Interaction, Mohamed bin Zayed University of Artificial Intelligence, Abu Dhabi, UAE (e-mail: jiayeli@link.cuhk.edu.cn; elizabeth.churchill@mbzuai.ac.ae).}% 
\thanks{T. Chen, S. Ma and K. Wu are with the Department of Robotics, Mohamed bin Zayed University of Artificial Intelligence, Abu Dhabi, UAE (e-mail: tongshun.chen@mbzuai.ac.ae; siyi.ma@mbzuai.ac.ae; ke.wu@mbzuai.ac.ae).}% 
}

% \author{\IEEEauthorblockN{1\textsuperscript{st} Jiaye Li}
% \IEEEauthorblockA{\textit{Mohamed bin Zayed University of} \\ \textit{Artificial Intelligence}\\
% Abu Dhabi, UAE \\
% jiayeli@link.cuhk.edu.cn}
% \and
% \IEEEauthorblockN{2\textsuperscript{nd} Tongshun Chen}
% \IEEEauthorblockA{\textit{Mohamed bin Zayed University of} \\ \textit{Artificial Intelligence}\\
% Abu Dhabi, UAE \\
% tongshun.chen@mbzuai.ac.ae}
% \and
% \IEEEauthorblockN{3\textsuperscript{rd} Siyi Ma}
% \IEEEauthorblockA{\textit{Hunan Agricultural University}\\
% Changsha, China \\
% 1183533766@stu.hunau.edu.cn}
% \and
% \IEEEauthorblockN{4\textsuperscript{th} Elizabeth Churchill}
% \IEEEauthorblockA{\textit{Mohamed bin Zayed University of} \\ \textit{Artificial Intelligence}\\
% Abu Dhabi, UAE \\
% elizabeth.churchill@mbzuai.ac.ae}
% \and
% \IEEEauthorblockN{5\textsuperscript{th} Ke Wu}
% \IEEEauthorblockA{\textit{Mohamed bin Zayed University of} \\ \textit{Artificial Intelligence}\\
% Abu Dhabi, UAE \\
% ke.wu@mbzuai.ac.ae}
% }

\markboth{ICRA WORKSHOP ON AI-DRIVEN SOFT ROBOTICS: INNOVATIONS, CHALLENGES, AND FUTURE DIRECTIONS, JUNE 2026}{}

\maketitle

\begin{abstract}
  We introduce PuppetAI, a modular soft robot interaction platform. This platform offers a scalable cable-driven actuation system and a customizable, puppet-inspired robot gesture framework, supporting a multitude of interaction gesture robot design formats. The platform comprises a four-layer decoupled software architecture that includes perceptual processing, affective modeling, motion scheduling, and low-level actuation. We also implemented an affective expression loop that connects human input to the robot platform by producing real-time emotional gestural responses to human vocal input. For our own designs, we have worked with nuanced gestures enacted by ``soft robots'' with enhanced dexterity and ``pleasant-to-touch'' plush exteriors. By reducing operational complexity and production costs while enhancing customizability, our work creates an adaptable and accessible foundation for future tactile-based expressive robot research. Our goal is to provide a platform that allows researchers to independently construct or refine highly specific gestures and movements performed by social robots. 
\end{abstract}

\begin{IEEEkeywords}
Expressive robotics, Interactive plush puppet robots, Cable-driven soft robotics, Modular design
\end{IEEEkeywords}



\begin{figure*}[t] % [t] 表示放在页面顶部
  \centerline{\includegraphics[width=0.90\textwidth]{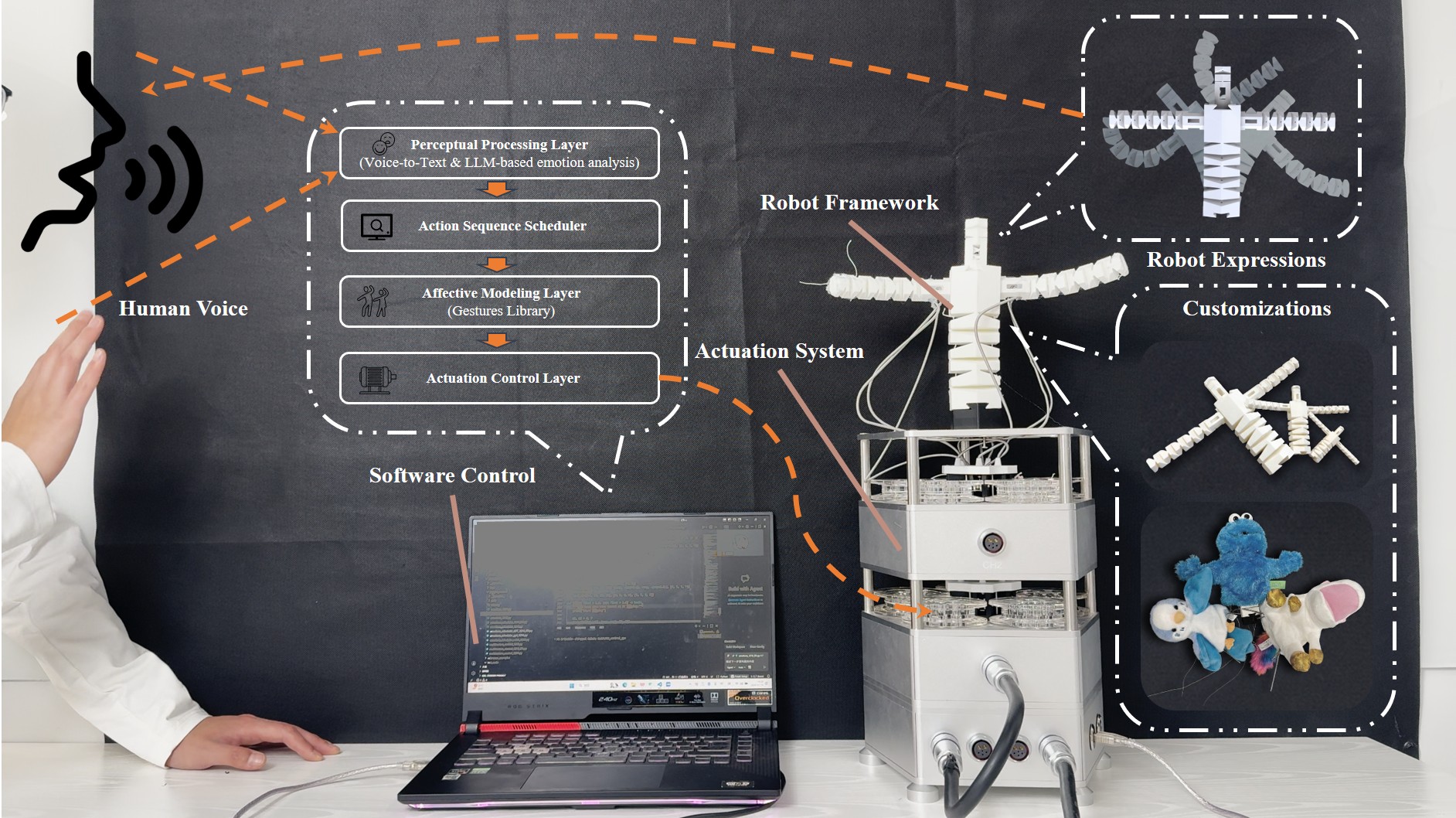}}
  \caption{Overview of the platform and the affective expression loop. The human voice is captured and analyzed by the control software with an integrated LLM, which generates parameterized motion sequences for the platform to perform affective gestural feedback. The robot’s gestures, in turn, shape subsequent human responses, completing the loop.}
  \label{fig:Interactive Loop}
\end{figure*}

% \begin{figure}[htbp]
%   \centering
%   \includegraphics[width=0.85\linewidth]{Figures/Loop2.jpg}
%   \caption{Overview of the interactive loop. Human vocal input is captured and analyzed by the control software with integrated LLM, which outputs parameterized motion sequences for the platform to perform affective gestural feedback. The robot’s gestures, in turn, shape subsequent human responses, completing the loop}
%   \label{fig:Interactive Loop}
% \end{figure}

\section{Introduction}
\label{sec:intro}
As robots become increasingly involved in our daily lives, Human-Robot Interaction (HRI) research, especially regarding expressive robots, has received significant attention \cite{saerbeck2010expressive, venture2019robot, mahadevan2024generative}. Puppet-inspired robots, due to their approachable aesthetic across all age groups and capacity for rich tactile engagement \cite{macari2021puppets, tai2011touching, remer2015teach, jeong2018huggable}, represent an ideal embodiment for expressive robotic interaction. Researchers have developed expressive interaction methods for puppet robots \cite{sugiura2012pinoky, sakashita2017you, liu2019hinhrob, she2020robot, martinez2014emopuppet} and deployed them in educational and play settings to explore interaction scenarios and establish benchmarks for expressive behavior \cite{causo2015developing, gupta2014puppetx}. However, many existing studies still depend on rigid internal structures and offer limited customization capacity across different sizes and structures of puppets. In addition, the high costs and proprietary nature of some platforms create barriers for researchers seeking to extend existing work across different research settings.

% However, many existing studies remain constrained by the morphological rigidity of traditional platforms. Most expressive robots rely on fixed joint structures and hard materials, such as plastic or metal, which limit their gestural range and exclude the critical dimension of tactile engagement. In addition, the high cost and proprietary nature of many platforms create a barrier for researchers seeking to extend existing work across different research settings.

Compared to rigid structures, soft robots, especially cable-driven continuum robots, offer a more compliant and flexible motion paradigm while maintaining mechanical simplicity \cite{Dewi2024LightweightModular}. Actuators can be placed remotely, reducing moving mass and improving operational safety, which makes these robots suitable for human-centered research. Moreover, by routing cables through multiple serially connected segments, researchers can customize the morphological configuration and bending range to meet specific task requirements without increasing structural complexity \cite{Liu2023CableContinuumSurvey}.
Despite these advances, soft robotic systems are currently often introduced as standalone concepts rather than being applied as functional platforms for HRI studies. 

To bridge these gaps, we propose PuppetAI, a modular soft robot interaction platform consisting of a scalable actuation system and a customizable, puppet-inspired cable-driven continuum robot gesture framework, designed for adaptive and accessible HRI research. Building on the mechanism introduced by Wang et al. \cite{wang2025spirobs}, our framework supports greater dexterity in gesturing and can be easily adapted to various puppet sizes and styles. Our design also allows the robot to have interchangeable plush, ``pleasant-to-touch'' exteriors for future tactile-based affective studies. Moreover, the actuation system allows the number of motors to be scaled based on specific research requirements to optimize control efficiency and reduce operational complexity. 

The control software architecture of this platform includes four decoupled layers: a perceptual processing layer that handles voice-to-text conversion and large language model (LLM)-based emotion analysis; an affective modeling layer that organizes emotional response synthesis; an action sequence scheduler that manages motion sequences; and a low-level actuation control layer for motor execution. The modular structure ensures that each component can be independently refined or replaced, providing a flexible testing ground for the research community. Additional implementation details are described in Section \ref{subsec:affective pipeline}.

In order to validate the utility of the platform, we implemented an affective expression loop that links human users with the robot platform by producing real-time emotional gestural responses to human vocal input. The structure of the loop is illustrated in Fig. \ref{fig:Interactive Loop}. For the affective modeling method, we developed a library of expressive behaviors inspired by professional puppetry techniques. We prompted a ChatGPT-based LLM to generate and organize parameterized response sequences for the scheduler. Overall, by reducing operational complexity and production costs, this work aims to provide a foundation and broaden access for future HRI research, empowering smaller laboratories to explore multimodal affective interactions through a customizable and tactile-rich soft robotic embodiment.

\section{Related Work}
\label{sec:related work}

\begin{figure*}[t]
  \centering
  \includegraphics[width=\linewidth]{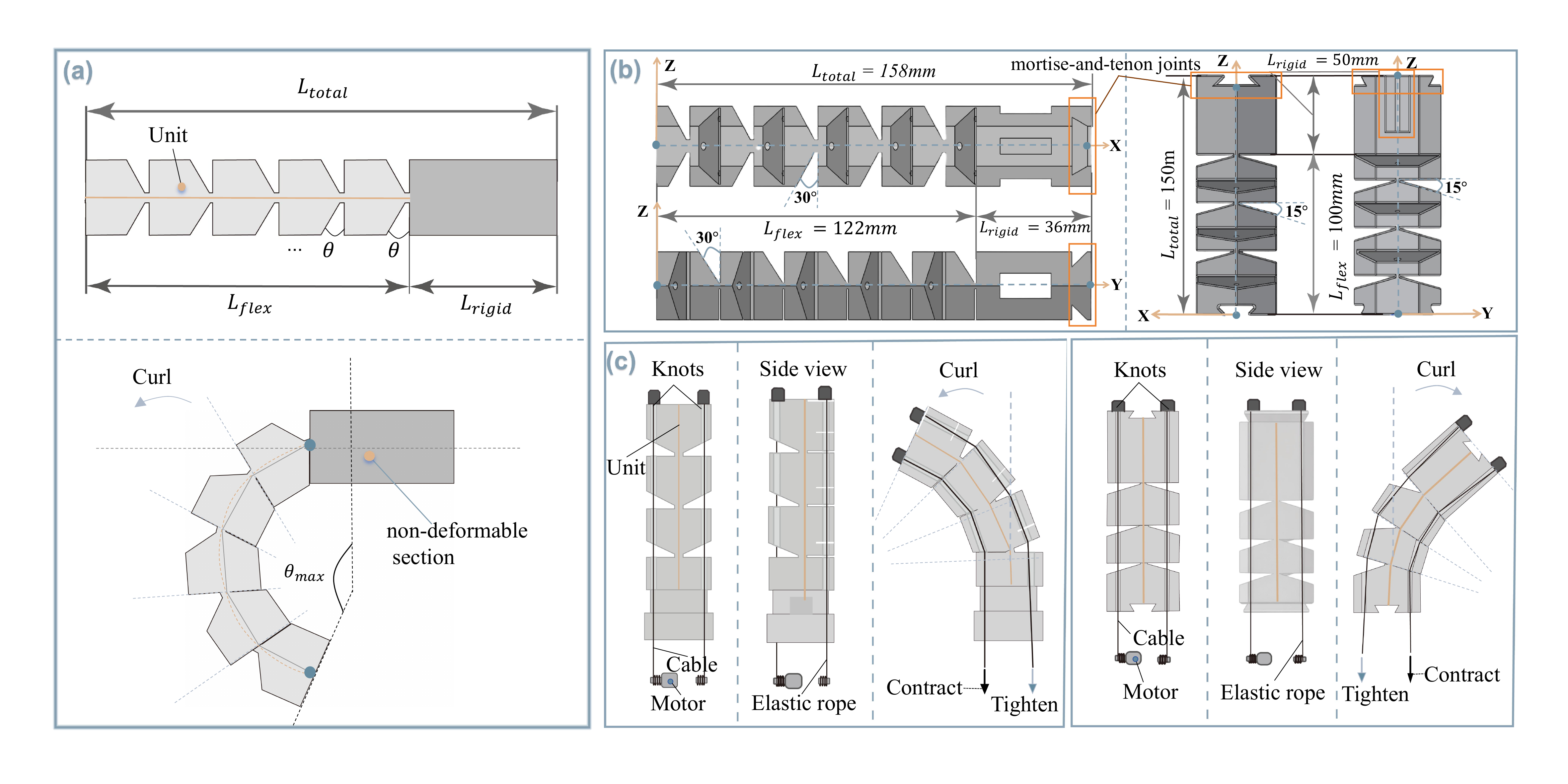}
  \caption{Overview of the continuum robot framework design. The robot’s structure consists of deformable and non-deformable sections, with the bending range adjusted by dividing the flexible part into specific units. It is powered by base-mounted motors and cables, using elastic ropes to provide a restoring force and simplify the control of planar motions.}
  \label{fig1}
\end{figure*}

\subsection{Robotic Puppetry}
A puppet is a movable, inanimate object manipulated by a puppeteer to convey expressive behaviors. Through body movements, such as motions of the hands and head, a puppet can communicate emotions and intentions visually \cite{kroger2019puppet}. 
% A puppeteer may further animate the puppet by providing a voice. Through the integration of movement and speech, an inanimate object can appear to come to life in the hands of a puppeteer \cite{kroger2019puppet}.

Prior research has explored various mechanical and computational methods for robotic puppetry. HinHRob was designed to perform glove puppetry using a robotic arm and hand with six and seven degrees of freedom \cite{she2020robot, liu2019hinhrob}. PINOKY is a wireless, ring-like device that can be attached to plush toys to animate their limbs \cite{sugiura2012pinoky}. Sakashita et al. \cite{sakashita2017you} presented a telepresence puppetry system that maps a human performer’s body and facial movements onto a puppet while providing audiovisual feedback to the performer. 
The application of puppet robots has also been studied in specific contextual environments, such as education and play. Causo et al. \cite{causo2015developing} developed robotic puppets to support teaching in kindergarten education, while PuppetX \cite{gupta2014puppetx} proposed a framework for both constructing playthings and playing with them using spatial body and hand gestures.
Besides academic research, industrial practices have produced remotely controlled puppet robots capable of subtle head and mouth movements for commercial applications \cite{Axtell2020}. However, many existing approaches are often characterized by specialized manipulation mechanisms with limited morphological flexibility. These systems lack the structural adaptability to accommodate puppets of varying sizes and structures. In addition, some platforms also employ complex or closed-source designs, which limit reproducibility and make it difficult for future researchers to build upon their work.

\subsection{Cable-driven continuum robots}

Cable-driven continuum robots offer a compliant and flexible motion paradigm while preserving mechanical simplicity and scalability. By regulating cable tensions, smooth bending and shape deformation can be achieved without rigid joints, resulting in inherent underactuation \cite{Dewi2024LightweightModular}. Moreover, actuators can be placed remotely, reducing moving mass and improving operational safety, which makes cable-driven robots suitable for human-centered environments.

Continuum robots typically consist of multiple serially connected segments, each with constrained bending capabilities. By adjusting the number of segments and the bending range of each segment, the overall robot configuration can be customized to meet different task requirements \cite{Dewi2024LightweightModular,Liu2023CableContinuumSurvey}. This modular design improves adaptability. Cable-driven actuation is also well suited for such modular designs, as cables can be routed through multiple segments to achieve coordinated deformation while maintaining structural simplicity \cite{Liu2023CableContinuumSurvey}. Despite these advantages, existing research on cable-driven continuum robots has mainly focused on mechanical modeling, motion control, and manipulation tasks. Their potential for emotion-oriented HRI studies remains largely unexplored. 

\section{Design of the PuppetAI}
\label{sec:design}

% \subsection{[name] and Affective Pipeline Structure}
PuppetAI is a modular, soft robot interaction platform that includes a scalable actuation system and a customizable, puppet-inspired cable-driven continuum robot gesture framework. This platform features a plush puppet body housing the internal soft framework, which is mounted on the actuation base. Driven by the actuation system, the puppet can produce a wide range of expressive gestures. Section \ref{subsec:platform} explains the design and fabrication of the robot framework; Section \ref{subsec:affective pipeline} outlines the software control architecture of the platform.

\subsection{Cable-driven Continuum Robot Gesture Framework}
\label{subsec:platform}

The platform employs a cable-driven continuum robot to enable compliant and expressive motion for interactive applications. It consists of a modular actuation base and a TPU-95 soft robot framework. This design allows the number of motors, cable routing, and structural dimensions to be reconfigured for different degrees of freedom (DoF) and robot sizes. For better demonstration, the currently implemented robot features a standard puppet morphology, consisting of a torso and two arms, both of which support spatial motion with two orthogonal bending planes separated by $90^\circ$. Non-deformable sections of the arms and body are connected via mortise-and-tenon joints.

As illustrated in Fig.~\ref{fig1}(a), the body consists of a deformable section of length $L_{\text{flex}}$ and a non-deformable section of length $L_{\text{rigid}}$, 
with the total length $L_{\text{total}} = L_{\text{flex}} + L_{\text{rigid}}$. 
The maximum bending angle $\theta_{\max}$ is achieved by dividing the deformable section into $n$ deformation units with equal 
cutting angles $\theta = \theta_{\max}/n$. By adjusting $L_{\text{flex}}$ and $n$, the robot scale and deformation range can be adapted to different puppet embodiments.

In Fig.~\ref{fig1}(b), each arm includes a 122\,mm deformable section and a 36\,mm non-deformable section, enabling vertical and forward swinging motions up to 150$^\circ$, implemented using five cutting units with 30$^\circ$ per unit. The body features a 100\,mm deformable section and supports forward, backward, and lateral bending up to 45$^\circ$, realized using three cutting units with 15$^\circ$ per unit.

As demonstrated in Fig.~\ref{fig1}(c), the robot is actuated by base-mounted motors that drive cables to generate bending motions. The three-dimensional motion of each driving module is decomposed into two planar motions, corresponding to two orthogonal bending planes. Within each plane, bending is produced by a single motor-driven cable paired with an elastic rope on the opposite side, providing a passive restoring force and reducing active control inputs.

\renewcommand{\arraystretch}{1.5}
\begin{table*}[t]
\caption{Library of Robot Affective Gestures. Motions are categorized as either Discrete (finite action units) or Continuous (sustained states). The neutral position is defined as both arms extended laterally, slightly below shoulder height.}
\begin{center}
\begin{tabular}{|l|l|p{10cm}|}
\hline
\textbf{Gesture} & \textbf{Type} & \textbf{Kinematic Description} \\
\hline
\textit{Waving} & Discrete & Unilateral arm elevation and depression; repeated in a rapid cycle. \\
\hline
\textit{Joy} & Discrete & Bilateral arm elevation performed simultaneously with a brief hold. \\
\hline
\textit{Sadness} & Discrete & Bilateral arm depression performed simultaneously with a brief hold. \\
\hline
\textit{Hug} & Discrete & Bilateral arm flexion inward toward the torso to form an embrace gesture; held briefly before releasing. \\
\hline
\textit{Confusion} & Discrete & Lateral torso lean combined with unilateral arm flexion toward the head to form a scratching pose; held briefly before releasing. \\
\hline
\textit{Dancing} & Continuous & Alternating unilateral arm flexion toward the torso; synchronized with rhythmic body oscillations. \\
\hline
\end{tabular}
\label{tab:motion_library}
\end{center}
\end{table*}

\subsection{Software Control Architecture}
\label{subsec:affective pipeline}
The software control is hosted on a dedicated computer and is interfaced with the platform via a USB-B cable. The software control architecture contains four modular layers: (1) a perceptual processing layer for voice-to-text conversion and large language model (LLM)-based emotion analysis; (2) an affective modeling layer for emotional response synthesis; (3) an action sequence scheduler to manage motion sequences; and (4) a low-level actuation control layer for motor execution. The low-level actuation layer manages motor operations, including phase-shifted multi-motor channel control, sinusoidal displacement, and torque limit detection. Building on this foundation, the affective modeling layer organizes motor commands to generate expressive robot gestures that correspond to different emotional states. Further details regarding the expressive motion design can be found in Section \ref{subsec:expressive motions design}. This layer is also designed to support the integration of new expressions.

For the perceptual processing layer, we employ the SenseVoiceSmall model to extract both textual transcriptions and vocal emotion estimates from user input. These data are then processed by a prompted ChatGPT-based LLM, which synthesizes action sequences by retrieving motions from a predefined gesture library hosted within the affective modeling layer. Sequences follow a structured format (e.g., [Waving][2][Happy][2]...), where numerical values indicate either the interval between instantaneous actions or the duration of continuous motions. These generated action sequences are executed by the action sequence scheduler to control robot motions.

\section{Interactive Loop}
\label{sec:application}

Based on the platform, we implemented an affective expression loop to demonstrate its utility. This loop mediates real-time emotional gestural responses to vocal input between human users and the robot platform, as shown in Fig. \ref{fig:Interactive Loop}. The robot's structural design facilitates efficient adaptation to a wide range of doll morphologies and sizes (see Customization in Fig. \ref{fig:Interactive Loop}). Our methodology for designing the affective expressions of the robot is described in Section \ref{subsec:expressive motions design}.

\subsection{Expressive Gestures Design}
\label{subsec:expressive motions design}
To maintain the semantic fidelity of emotional expression, two researchers conducted open coding on video and text archives of professional puppet performances. Based on our findings, we designed an expressive motion library for the affective modeling layer. We categorize these motions into two distinct types: discrete gestures, defined as finite, self-contained action units (e.g., waving, hugging, and joy), and continuous gestures, which require sustained actions to convey a state (e.g., dancing). By grounding these motions in traditional performance practices, we aim to ensure the robot conveys the intended emotional states. 

For demonstration, the designed affective gestures were implemented on a standard puppet morphology explained in Section \ref{subsec:platform}. Table \ref{tab:motion_library} details the kinematic execution of some selected motions relative to the robot's neutral ``idle'' state, where both arms extend laterally, slightly below shoulder height. When not actively responding to user input, the robot performs subtle, randomized lateral swaying to maintain a sense of liveliness.

\subsection{Interaction Scenarios}
The robot performs expressive motions in response to the semantic and emotional content of the user's speech. The current behavioral model focuses on empathetic resonance rather than active emotion regulation. For instance, the robot responds to positive user affect with celebratory motions and to negative affect with sympathetic gestures. The following scenarios illustrate representative interactions:

% \begin{table}[htbp]
%   \caption{Representative Interaction Scenarios. Each scenario represents a specific human vocal input to a parameterized sequence of affective gestures executed by the platform.}
%   \label{tab:interaction_scenarios}
%   \centering
%   \begin{tabular}{l p{5cm} p{6cm}}
%     \toprule
%     \textbf{Scenario} & \textbf{Human Input} & \textbf{Robot Response (Gesture Sequence \& Behavior)} \\
%     \midrule
%     \textit{Greeting} & ``Hi, how are you today?'' & \texttt{[Waving][1][Joy][1]} Waving gesture followed by raising its hands to express joy. \\
    
%     \textit{Positive} & ``Wow, I had a really wonderful day!'' & \texttt{[Joy][1][Dancing][3]} Raising arms in delight and transitioning into a synchronized dance sequence. \\
    
%     \textit{Negative} & ``I stayed up late to work yesterday; I am very tired.'' & \texttt{[Sadness][1][Hug][3]} Lowering arms to indicate sadness, followed by a comforting embrace gesture. \\
    
%     \textit{Confusion} & ``Who are you?'' & \texttt{[Confused][1]} Leaning forward with bent arms and a questioning gesture to convey puzzlement. \\
%     \bottomrule
%   \end{tabular}
% \end{table}

\begin{itemize}
    \item \textbf{Greeting:} 
    \textit{Human:} ``Hi, how are you today?'' \\
    \textit{Example Response:} \texttt{[Waving][1][Joy][1]} \\
    The robot performs a waving gesture followed by raising its hands to express joy.
    
    \item \textbf{Positive Feeling:} 
    \textit{Human:} ``Wow, I had a really wonderful day!'' \\
    \textit{Example Response:} \texttt{[Joy][1][Dancing][3]} \\
    The robot raises its arms in delight and transitions into a dance sequence.
    
    \item \textbf{Negative Feeling:} 
    \textit{Human:} ``I stayed up late to work yesterday; I am very tired.'' \\
    \textit{Example Response:} \texttt{[Sadness][1][Hug][3]} \\
    The robot lowers its arms to indicate sadness, followed by a comforting hug gesture.
    
    \item \textbf{Confusion:} 
    \textit{Human:} ``Take a guess at how many candies I have in my pocket.'' \\
    \textit{Example Response:} \texttt{[Confusion][1]} \\
    The robot leans forward with bent arms to form a questioning gesture.
\end{itemize}

Voice interaction is currently triggered through a discrete push-to-talk mechanism (button press to initiate and terminate recording). Future iterations will incorporate Voice Activity Detection and Keyword Spotting to enable hands-free conversational turn-taking.

% \subsection{Size and Appearance Customization}
% We used a diverse selection of commercially available puppets sourced from online marketplaces to serve as the robot's soft exterior. These puppets were not specifically designed for this hardware and feature varying lengths and widths in their bodies and limbs. Our architecture and platform are designed to adapt effectively to these differences. We integrated a modular mortise-and-tenon mechanism to adjust the arm and torso lengths. Meanwhile kinematic topology and control algorithms will not be affected by these changes. Figure \ref{fig:supported_puppets} illustrates some examples of supported morphologies. 

% \input{Sections/Challenges and Future Directions}

\section{Conclusion}
\label{sec:conclusion}
A key aspect of interaction is a feeling of positive connection. Most robots, even those designed for connection, rely on rigid structures and thus constrain their potential to inspire human-robot connection. Our work provides a modular interaction platform for ``soft, emotionally connected robots'' designed to support tactile-based expressive robotics research. By combining a scalable actuation system with a customizable, puppet-inspired, cable-driven continuum robot gesture framework, our platform allows for increased flexibility of interaction and overcomes the morphological rigidity and limited adaptability of traditional expressive robots. This design supports enhanced gestural dexterity and introduces a wider dimensionality of tactile expressiveness. Furthermore, the decoupled software architecture provides a flexible testing ground for researchers to independently isolate and refine specific gestures and actions. We implemented an affective expression loop featuring LLM-based emotion analysis and puppetry-inspired gesture design to demonstrate the platform's interaction capabilities. In the future, we aim to advance this platform into a fully integrated, standalone unit featuring real-time generative expression and expanded multimodal capabilities, such as subdermal pneumatic tactile systems and spatialized synthetic speech. By reducing operational complexity and production costs while increasing customizability, this work provides an adaptable and accessible foundation for future tactile-based expressive robotics studies.

% \section{Appendices}

\bibliographystyle{IEEEtran}
\bibliography{sample-base}

\begin{thebibliography}{10}
\providecommand{\url}[1]{#1}
\csname url@rmstyle\endcsname
\providecommand{\newblock}{\relax}
\providecommand{\bibinfo}[2]{#2}
\providecommand\BIBentrySTDinterwordspacing{\spaceskip=0pt\relax}
\providecommand\BIBentryALTinterwordstretchfactor{4}
\providecommand\BIBentryALTinterwordspacing{\spaceskip=\fontdimen2\font plus
\BIBentryALTinterwordstretchfactor\fontdimen3\font minus \fontdimen4\font\relax}
\providecommand\BIBforeignlanguage[2]{{%
\expandafter\ifx\csname l@#1\endcsname\relax
\typeout{** WARNING: IEEEtran.bst: No hyphenation pattern has been}%
\typeout{** loaded for the language `#1'. Using the pattern for}%
\typeout{** the default language instead.}%
\else
\language=\csname l@#1\endcsname
\fi
#2}}

\bibitem{saerbeck2010expressive}
M.~Saerbeck, T.~Schut, C.~Bartneck, and M.~D. Janse, ``Expressive robots in education: varying the degree of social supportive behavior of a robotic tutor,'' in \emph{Proceedings of the SIGCHI conference on human factors in computing systems}, 2010, pp. 1613--1622.

\bibitem{venture2019robot}
G.~Venture and D.~Kuli{\'c}, ``Robot expressive motions: a survey of generation and evaluation methods,'' \emph{ACM Transactions on Human-Robot Interaction (THRI)}, vol.~8, no.~4, pp. 1--17, 2019.

\bibitem{mahadevan2024generative}
K.~Mahadevan, J.~Chien, N.~Brown, Z.~Xu, C.~Parada, F.~Xia, A.~Zeng, L.~Takayama, and D.~Sadigh, ``Generative expressive robot behaviors using large language models,'' in \emph{Proceedings of the 2024 ACM/IEEE International Conference on Human-Robot Interaction}, 2024, pp. 482--491.

\bibitem{macari2021puppets}
S.~Macari, X.~Chen, L.~Brunissen, E.~Yhang, E.~Brennan-Wydra, A.~Vernetti, F.~Volkmar, J.~Chang, and K.~Chawarska, ``Puppets facilitate attention to social cues in children with asd,'' \emph{Autism Research}, vol.~14, no.~9, pp. 1975--1985, 2021.

\bibitem{tai2011touching}
K.~Tai, X.~Zheng, and J.~Narayanan, ``Touching a teddy bear mitigates negative effects of social exclusion to increase prosocial behavior,'' \emph{Social Psychological and Personality Science}, vol.~2, no.~6, pp. 618--626, 2011.

\bibitem{remer2015teach}
R.~Remer and D.~Tzuriel, ``“i teach better with the puppet”--use of puppet as a mediating tool in kindergarten education--an evaluation,'' \emph{American Journal of Educational Research}, vol.~3, no.~3, pp. 356--365, 2015.

\bibitem{jeong2018huggable}
S.~Jeong, C.~Breazeal, D.~Logan, and P.~Weinstock, ``Huggable: the impact of embodiment on promoting socio-emotional interactions for young pediatric inpatients,'' in \emph{Proceedings of the 2018 CHI conference on human factors in computing systems}, 2018, pp. 1--13.

\bibitem{sugiura2012pinoky}
Y.~Sugiura, C.~Lee, M.~Ogata, A.~Withana, Y.~Makino, D.~Sakamoto, M.~Inami, and T.~Igarashi, ``Pinoky: a ring that animates your plush toys,'' in \emph{Proceedings of the SIGCHI Conference on Human Factors in Computing Systems}, 2012, pp. 725--734.

\bibitem{sakashita2017you}
M.~Sakashita, T.~Minagawa, A.~Koike, I.~Suzuki, K.~Kawahara, and Y.~Ochiai, ``You as a puppet: evaluation of telepresence user interface for puppetry,'' in \emph{Proceedings of the 30th annual ACM symposium on user Interface software and technology}, 2017, pp. 217--228.

\bibitem{liu2019hinhrob}
H.~Liu, Y.~She, L.~Lin, S.~Chen, J.~Chen, X.~Xu, and J.~Lin, ``Hinhrob: A performance robot for glove puppetry,'' in \emph{SIGGRAPH Asia 2019 Posters}, 2019, pp. 1--2.

\bibitem{she2020robot}
Y.~She, X.~Xu, H.~Liu, J.~Lin, M.~Yang, L.~Lin, and B.~Yang, ``A robot for interactive glove puppetry performance,'' in \emph{International Conference on Computer Animation and Social Agents}.\hskip 1em plus 0.5em minus 0.4em\relax Springer, 2020, pp. 31--40.

\bibitem{martinez2014emopuppet}
J.~I. Mart{\'\i}nez, ``emopuppet: Low-cost interactive digital-physical puppets with emotional expression,'' in \emph{Proceedings of the 11th conference on advances in computer entertainment technology}, 2014, pp. 1--4.

\bibitem{causo2015developing}
A.~Causo, G.~T. Vo, E.~Toh, I.-M. Chen, S.~H. Yeo, and P.~W. Tzuo, ``Developing and benchmarking show \& tell robotic puppet for preschool education,'' in \emph{2015 IEEE International Conference on Robotics and Automation (ICRA)}.\hskip 1em plus 0.5em minus 0.4em\relax IEEE, 2015, pp. 6114--6119.

\bibitem{gupta2014puppetx}
S.~Gupta, S.~Jang, and K.~Ramani, ``Puppetx: A framework for gestural interactions with user constructed playthings,'' in \emph{proceedings of the 2014 international working conference on advanced visual interfaces}, 2014, pp. 73--80.

\bibitem{Dewi2024LightweightModular}
P.~T. Dewi, P.~Rao, and J.~Burgner-Kahrs, ``{A Lightweight Modular Segment Design for Tendon-Driven Continuum Robots with Pre-Programmable Stiffness},'' in \emph{2024 IEEE International Conference on Soft Robotics (RoboSoft)}.\hskip 1em plus 0.5em minus 0.4em\relax IEEE, 2024, pp. 531--536.

\bibitem{Liu2023CableContinuumSurvey}
Y.~Liu, K.~Zhang, B.~Huo, and P.~Chen, ``{A Review on Status and Prospects of Tendon/Cable Driven Continuum Robot},'' \emph{Journal of Zhengzhou University (Engineering Science)}, vol.~44, no.~3, pp. 1--11, 2023.

\bibitem{wang2025spirobs}
Z.~Wang, N.~M. Freris, and X.~Wei, ``Spirobs: Logarithmic spiral-shaped robots for versatile grasping across scales,'' \emph{Device}, vol.~3, no.~4, 2025.

\bibitem{kroger2019puppet}
T.~Kr{\"o}ger and A.-M. Nupponen, ``Puppet as a pedagogical tool: a literature review.'' \emph{International electronic journal of elementary education}, vol.~11, no.~4, pp. 393--401, 2019.

\bibitem{Axtell2020}
\BIBentryALTinterwordspacing
A.~Expressions. (2020) Easy-talk | fun. simple. affordable! [Online]. Available: \url{https://axtell.com/easy-talk/}
\BIBentrySTDinterwordspacing

\end{thebibliography}

% \appendix

\end{document}